\documentstyle[pre,aps,tighten,epsf,floats]{revtex}
\begin{document}
\title{Multiple current reversals in forced inhomogeneous ratchets} 
\author{\large Debasis Dan,$^{1,*}$ Mangal C. Mahato,$^{2}$ andA. M. Jayannavar$^{1,\dagger }$ \vspace{0.2in}}\address{\large $^1$ Institute of Physics, Sachivalaya Marg,
Bhubaneswar 751005, India}
\address{\large $^2$ Department of Physics, Guru Ghasidas University, Bilaspur 495009, India}\maketitle
\maketitle
\begin{abstract}
  
  Transport properties of overdamped Brownian paricles in a rocked 
thermal ratchet with space dependent
friction coefficient is studied. By tuning the parameters, the
direction of current exhibit multiple reversals, both as a function
of the thermal noise strength as well as the amplitude of rocking
force. Current reversals also occur under
deterministic conditions and exhibits intriguing structure. All these
results arise due to mutual interplay between potential asymmetry,
noise, driving frequency and inhomogeneous friction.

\end{abstract}
\vspace{1.0in}
      Fluctuation induced transport in ratchet systems has been an 
active field of research over the last decade. In these systems in the
absence of any net microscopic forces, asymmetric potential can be 
used to induce a unidirectional particle flow when subjected to an external
nonthermal fluctuations. These studies have been motivated in 
part by the attempt
to understand the mechanism of movement of protein motors in biological 
systems~\cite{Motor}. To this effect several physical  models have been proposed under the 
name of rocking ratchets~\cite{Rock,Rock2,Main}, flashing
ratchets~\cite{Flash,Flash2},
diffusion ratchets~\cite{Diffu}, correlation
ratchets~\cite{Corr}, etc. In all these studies the potential is taken 
to be asymmetric
in space. It has also been shown that one can  obtain  unidirectional 
current 
in the presence of spatially symmetric potentials. For these nonequilibrium 
systems external random force should be time asymmetric~\cite{Ma}  or the presence
of space dependent mobility is
required~\cite{IntMod,Mill,Jay,But,Kam,Spc_mob,Dan3}. By suitably tuning the system 
parameters such as temperature, friction coefficient, mass, etc, one can 
even change the direction of the current. Indeed, the study of current 
reversal phenomena has given rise to a research activity on its own. 
The motivation  being the possibility of new particle separation
devices superior to existing methods  such as electrophoretic
method for particles of micrometer scale~\cite{Kep}. 

      Bartussek \textit{et. al.}~\cite{Main} showed the occurrence of
current reversal in
rocked thermal ratchet with both amplitude of rocking force as well as 
the temperature of thermal bath.  They attributed this current reversal to the 
``mutual interplay between noise and finite-frequency driving''.
Multiple current reversals
have also been shown in the deterministic limit of these ratchets when
the inertial term is taken into account~\cite{Rock2,Under}. However 
these multiple current reversals in inertial ratchets are not robust 
in the presence of noise. Beside rocking ratchets 
current reversals have also been observed in flashing
ratchets~\cite{Reim,Bier,Flash2}. We in this work report that
that multiple reversals can be achieved even in  rocked \textit{overdamped} ratchet
in the presence of space dependent mobility, as a function of the noise strength
and amplitude of the rocking force. More over these systems 
show current reversals when rocked adiabatically.
In the deterministic overdamped case we get current reversal as a function 
of the amplitude of rocking force. Most of our results are attributed to the 
presence of space dependent mobility.
We have studied the same system as due to Bartussek \textit{et
  al}~\cite{Main}, except the presence 
of space dependent friction term. In earlier work the spatial asymmetry of 
the potential is responsible for unidirectional currents and their reversal 
as function of frequency.  As mentioned previouly one can get  
unidirectional currents in the presence of symmetric potentials, but
for this space dependent friction is
required~\cite{IntMod,Mill,Jay,But,Kam,Spc_mob,Dan3}. In these systems transport
direction is given by the phase shift between mobility and periodic
potential. This phase shift induces spatial symmetry breaking as required for 
the directed motion.
Appropriately choosing the phase shift leads to current reversal. We would like to 
emphasize that space dependent friction does not alter the equilibrium 
property of the system, however when the system is driven out of equilibrium 
not trivial dynamical effects arise due to space dependent friction~\cite{Dan3,PLA,PRE}. Naturally
in our present system we expect additional effects arising due to combination 
of spatial asymmetry and position dependence of friction coefficient.
It is to be noted that systems with space dependent friction are not uncommon.
Brownian motion in confined geometries show space dependent friction~\cite{Conf}. 
Particles diffusing close to surface have space dependent 
friction coefficient~\cite{Conf,Surf}. It is believed that molecular motor 
proteins move 
close along the periodic structure of microtubules and will therefore 
experience a position dependent mobility~\cite{Spc_mob}. Frictional inhomogeneities are 
common in super lattice structures and Josephson
junctions~\cite{Falco} also.

          We consider an overdamped Brownian particle moving in a
asymmetric potential V(x) with space dependent friction coefficient 
$\eta$(x) under the influence of external force field $F$(t) at
temperature $T$. 
Throughout our analysis we take the ratchet potential
V(x) = $-\frac{1}{2 \pi} (\sin (2\pi x) + \frac{\mu}{4} \sin (4 \pi
x))$. Here $\mu$ is the asymmetry parameter with values taken in the
range $0 <\mu <1 $,  friction coefficient $\eta (x) = \eta_{0}(1-\lambda \sin
(2\pi x + \phi))$, $|\lambda| < 1$. $\phi$ determines the relative
phase shift between  friction coefficient and potential . 
The forcing term is 
taken to be F(t) = A$\sin (w t + \theta)$, ($w = \frac{2
  \pi}{\tau}$, where $\tau$ is period of force) . Without any 
loss of  generality $\theta$ is taken to be zero. 
 The correct Langevin equation for this system in the overdamped limit
is given by~\cite{IntMod,Pram,Sancho} 

\begin{equation}
 \dot{x} =  -\frac{(V'(x) - F(t))}{\eta (x)} - k_{B}T \frac{\eta '(x)}{(\eta 
   (x))^{2}} + \sqrt{\frac{k_{B}T}{\eta (x)}}\xi (t) ,
\label{Langvn}
\end{equation} 
where $\xi(t)$ is a zero mean thermal Gaussian noise with correlation 
$<\xi (t) \xi (t')> = 2 \delta (t-t')$
  The equation of motion is equivalently described by the Fokker
Planck Equation ( FPE)~\cite{IntMod} 

\begin{equation}
 \frac{\partial P(x,t)}{\partial t} = \frac{\partial}{\partial x}
 \frac{1}{\eta (x)} [k_{B}T \frac{\partial}{\partial x} + (V'(x) -
 F(t))] P(x,t) .
 \label{FPE}
\end{equation} 
where $P(x,t)$ is the probability density at position $x$ at time
$t$. Equation~(\ref{FPE}), in the form of a continuity equation 

\begin{equation}
 \frac{\partial P(x,t)}{\partial t} = - \frac{\partial
   J(x,t)}{\partial x} ,
 \label{Conti}
\end{equation}

where 
\begin{equation}
  J(x,t) = -\frac{1}{\eta(x)} [(V'(x)-F(t)) + k_{B}T 
  \frac{\partial}{\partial x}]P(x,t),
 \label{Current}
\end{equation}
is the probability current. Since the potential and the driving force
have spatial and temporal periodicity respectively, therefore $J(x,t)
= J(x + 1,t + \tau)$, ~\cite{Risk,Main}. The average current $j$
in the system is given by 
\begin{equation}
 j = \lim_{t \rightarrow \infty} \frac{1}{\tau} \int_{t}^{t+\tau} dt
 \int_{0}^{1} J(x,t) dx .
 \label{Avg_curr}
\end{equation}
It should be noted that for symmetric potential and $\lambda = 0$, $j
= 0$. Rectification of current is possible if the potential is either
asymmetric or $\lambda \neq 0$ with $\phi \neq 0, \pi$~\cite{Dan3}. $j$ is 
independent of the initial phase $\theta$ of the driving force.

 We explore various parameter regimes of the problem 
extensively by solving the FPE, Eq.~(\ref{FPE}) numerically with 
finite difference method. 
Throughout we have set current $j$ and all other physical quantities 
such as $T$, $A$, $w$ in dimensionless form .

\begin{figure}  
\centerline{\epsfbox{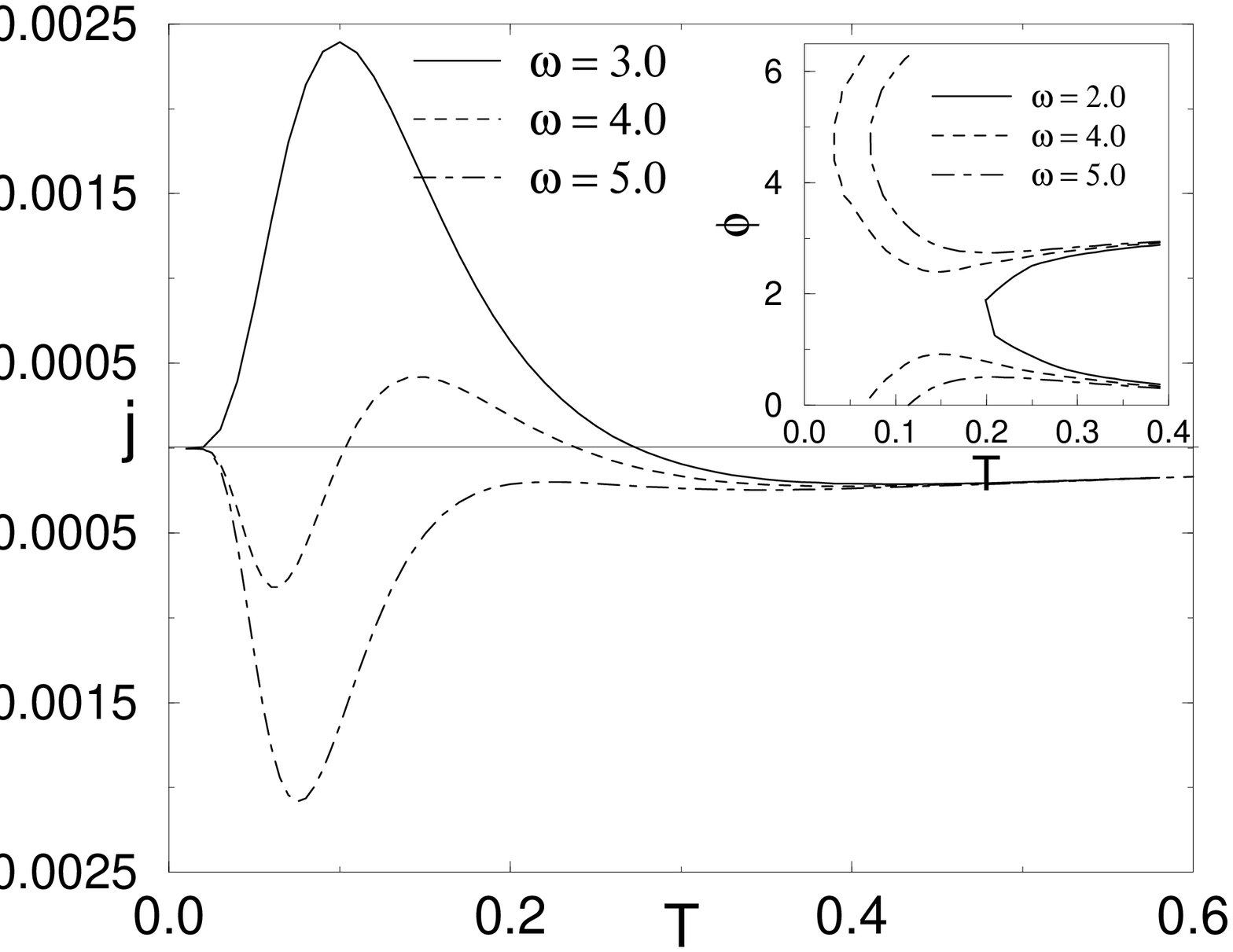}}
\vspace{.0in}
\caption{The mean current $j$ \textit{vs} temperature $T$ for $\phi =
  0.2 \pi$, $A=0.5$ and $\lambda = 0.1$. The driving frequencies are $w=
  3.0, 4.0$ and $5.0$. The inset shows the contour of zero current for 
  same values of $T, \lambda$. Regions enclosed on the right hand side 
  of a given contour is the negative current region and vice
  versa. Note that we have only one reversal for all values of $\phi > 
  \pi$.}
\label{j-T}
\end{figure}

\begin{figure}  
\centerline{\epsfbox{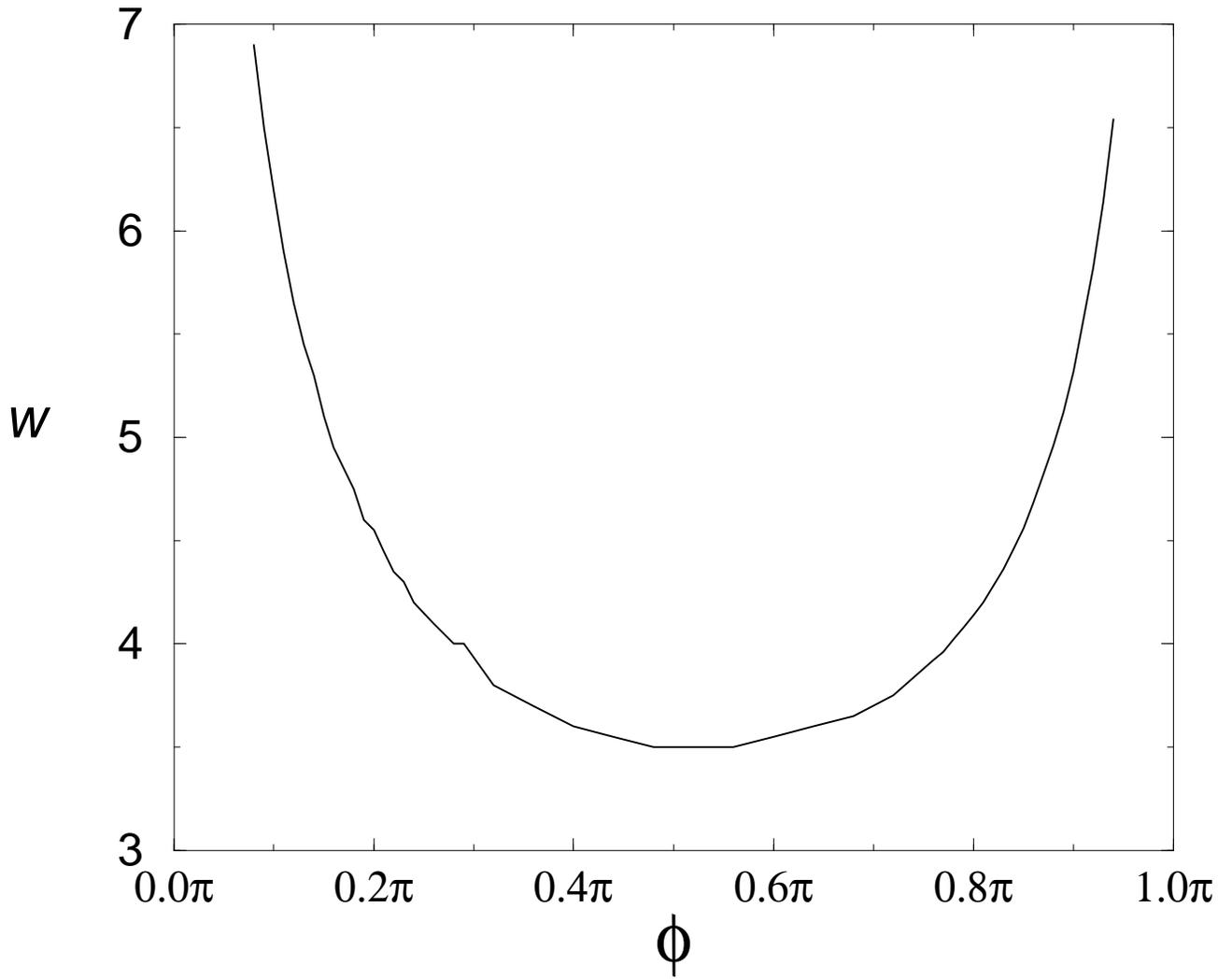}}
\caption{The critical $w$ above which no current reversal with
  $T$ occurs \textit{vs} $\phi$ for $A=0.5$ and $\lambda=0.1$.}
\label{wc_phi}
\end{figure}

\begin{figure}  
\centerline{\epsfbox{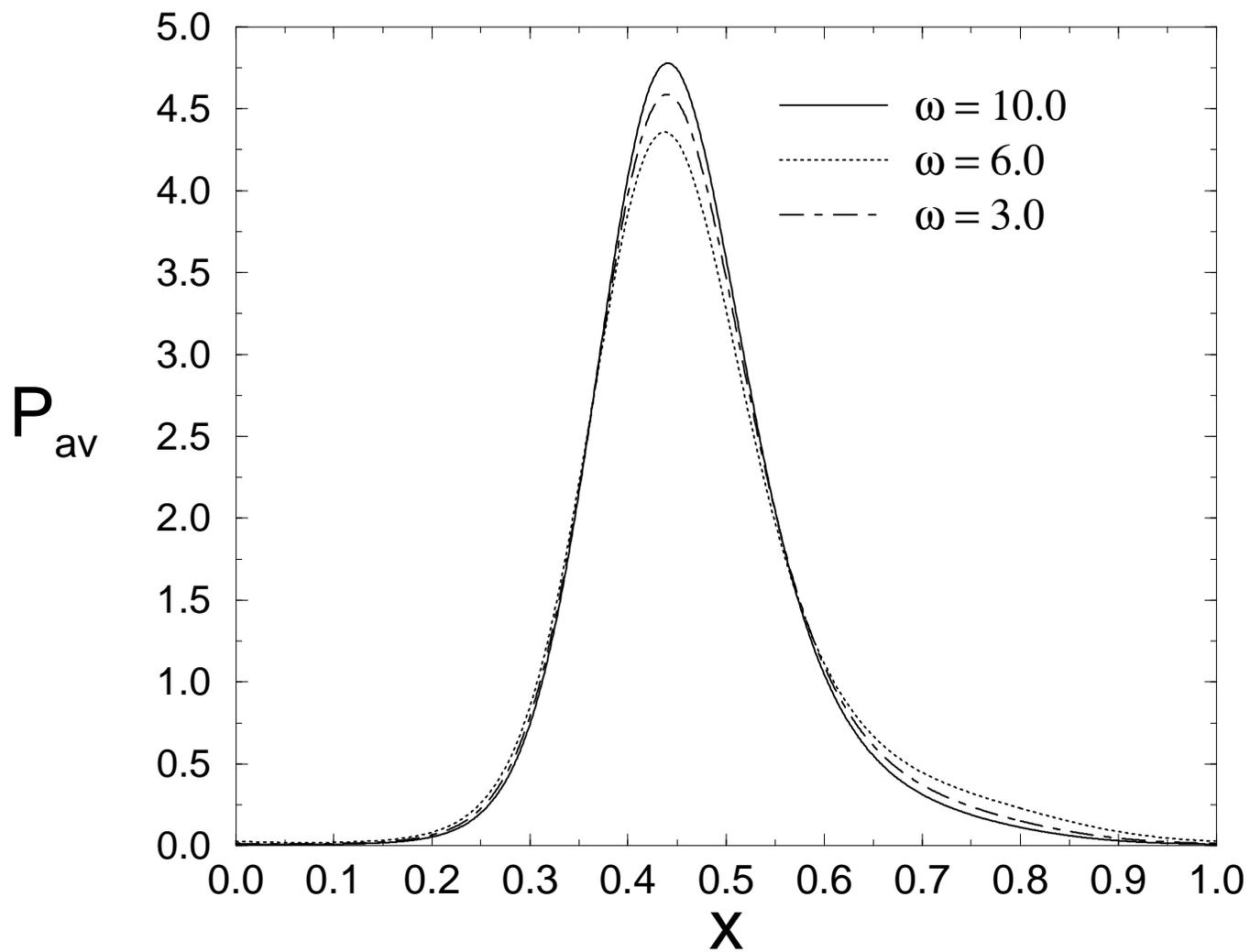}}
\vspace{.0in}
\caption{The time averaged probability $P_{av}$ \textit{vs} x for
  various values of $w$ and $T = 0.05$ and $A=0.5$. The
  distribution is independent of $\phi$ and $\lambda$.}
\label{P_av}
\end{figure}

\begin{figure} 
\centerline{\epsfbox{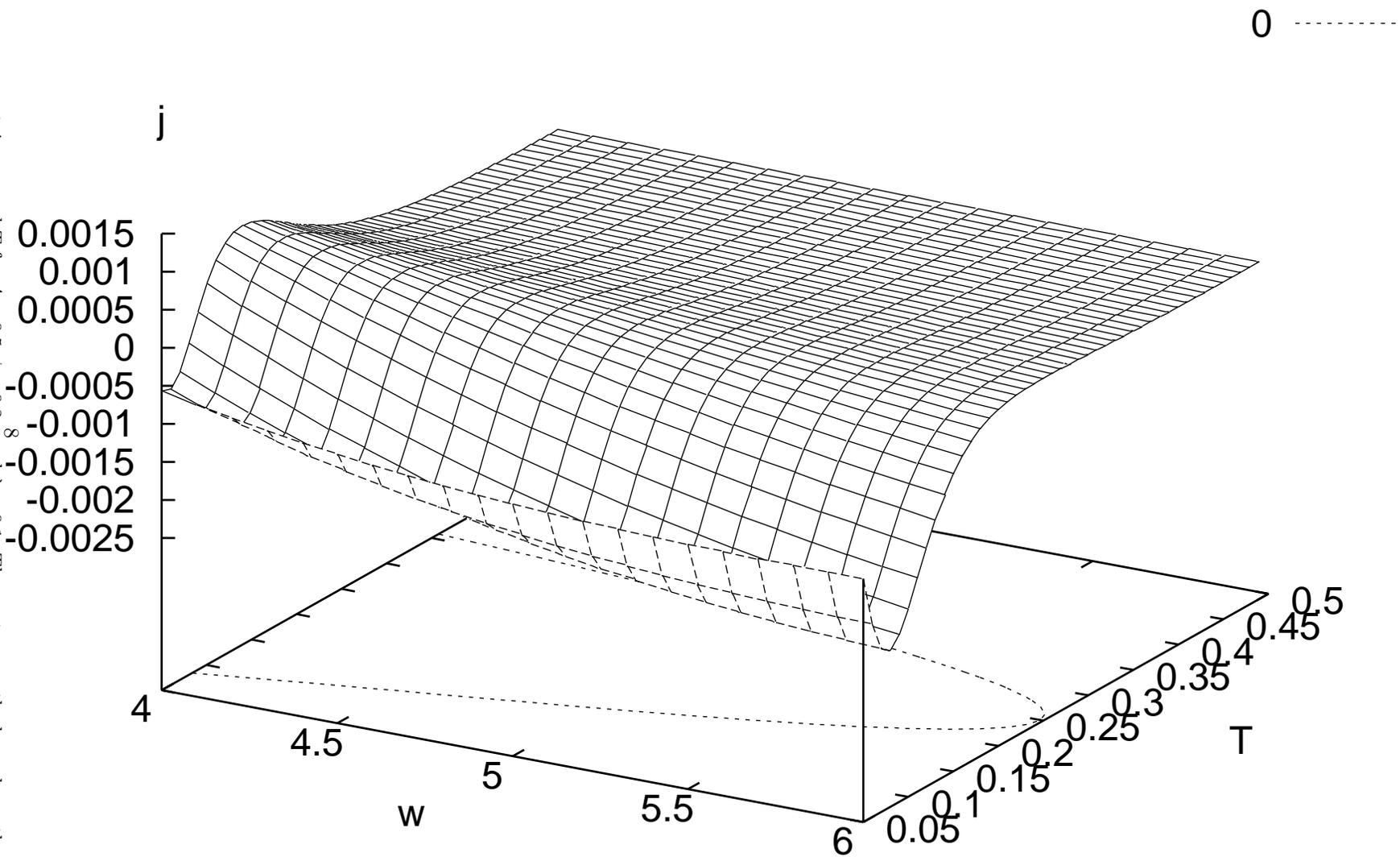}} 
\vspace{-1.1cm}
\caption{The mean current $j$ \textit{vs} $w$ and $T$ for $A = 0.5, 
  \phi=0.2 \pi$ and $\lambda = 0.1$. The contour on the base shows the 
  zero current.}
\label{j-T-w}
\end{figure}

 In the Fig.~\ref{j-T}, the average current $j$ is plotted 
\textit{vs} $T$ for various values of $w$. Here $\lambda = 0.1 , \phi =
0.2 \pi$, $A = 0.5$ and $\mu = 1$. Unlike Bartussek et al. ~\cite{Main}, 
where they show that
current reversal is not possible under adiabatic conditions,
however current reversal even in adiabatic condition can be obtained
in the present case for
some values of $\phi$~\cite{Dan3}. For moderately high frequency $w = 4.0$ 
the current reverses its sign \textit{twice} at $T=0.1$ and
$T=0.22$ and asymptotically goes to zero for higher values of $T$ as shown 
in the figure. This phenomena of twice current reversal with
temperature $T$ is the foremost feature of our system, previously    
unseen in any overdamped system. 
The inset shows the zero contour plot 
of the current $j$ versus $T$ and $\phi$ for three values of
$w$. Crossing this zero contour line implies current reversal. 
It can be seen that twice current reversal occurs only for a very narrow
range of $\phi$. For frequencies higher than certain 
critical frequency $w_{c}(\phi)$ current flows in only one direction for 
all temperatures, i.e no current reversal occurs. The plot
$w_{c}$ \textit{vs} $\phi$ is shown in Fig.~\ref{wc_phi}. 
The adiabatic curve  $w < 1$ is not shown in the figure as it
goes much beyond the scale of the graph. However it has a similar qualitative
shape as for $w = 3.0$ curve. As mentioned before, currents are 
due to the combined effect of phase shift $\phi$ coming from space dependent
friction and asymmetry parameter $\mu$. In the regime $\phi = 0.2 \pi$
and $A = 0.5 $ and in the absence of asymmetry current flows in the negative
direction for all values of $T$. The asymmetric case ( $\mu = 1.0$) in the
absence of space dependent friction give current in the positive direction
only as a function of temperature. Separately in both these cases 
absolute value of current
exhibits a maxima as a function of $T$, reminiscent of stochastic 
resonance phenomena. In a purely asymmetric case ($\lambda = 0$) current 
vanishes rapidly when $T$ exceeds the temperature associated with the 
barrier height. Whereas, in the symmetric case due to space dependent friction
absolute values of currents are significantly higher and decay slowly to 
zero in the large temperature regime. Naturally in the presence of 
both asymmetry and space dependent friction for the case under study 
the low temperature regime is dominated by the effect of asymmetry while the 
high temperature regime is dominated by space dependent friction. From this
, one can qualitatively explain the current reversals from positive 
to negative side as a function of temperature even in the adiabatic limit. 
In the regime $\phi > \pi$ current due to space dependent friction and 
potential asymmetry flows in same direction and hence no reversal is possible
as a function of $T$.  For
frequencies higher than the interwell frequency $w_{0}$, 
the low temperature scenario
changes. The direction of current in this regime is more of a
interplay between potential asymmetry and $w$ than that of
$\lambda$. Due to higher frequency the Brownian particles do not get
enough time to cross the right barrier which is at a larger distance 
from the minima. Number of particles moving about the 
potential minima increases. This fact is amply reflected in Fig.~\ref{P_av}, 
where the time averaged probability curve 
$ P_{av}(x) = \frac{1}{\tau} \int_{0}^{\tau} P(x,t) dt$
is plotted as function of $x$ for various values of
$w$ with $T = 0.05$. It is to be noted that the distribution is
independent of $\eta (x)$. Figure~\ref{P_av} shows that the 
probability of finding the
Brownian particles near the minima of the potential well increases with
increasing frequency, consequently the probable number of particles
near the potential barrier decreases. Since the distance from a
potential minima to the basin of attraction of next minima is less
from the steeper side than from the slanted side, hence in one period
the particles get enough time to climb the potential barrier from the
steeper side than from the slanted side, resulting in a negative
current. On increasing  the temperature , the particles get kicks of
larger intensity and hence they easily cross the slanted barrier,
resulting in a current reversal and positive current. On further
increasing the temperature, the effect of $\lambda$ dominates as
mentioned previously  and as a result the current
again flows in the negative direction, implying a second current
reversal as shown in dotted curve of Fig~\ref{j-T}. It is obvious from 
the above argument that for higher frequency the first reversal
temperature will be higher as shown in the Fig.~\ref{j-T-w}. The
dotted line in the base of Fig.~\ref{j-T-w} shows the contour of zero
current. But as mentioned previously, since the effect of $\lambda$ dominates 
for higher temperature, therefore the second reversal temperature
decreases with increasing frequency. Beyond the critical frequency $w_{c}$
there is no reversal of current as shown in
Fig.(~\ref{j-T-w},~\ref{wc_phi}). In the high frequency regime 
the effect of space dependent friction dominates the nature of current. 
We would like to emphasize here that in absence of asymmetry current reversals does
not take place. 

\begin{figure}
\centerline{\epsfbox{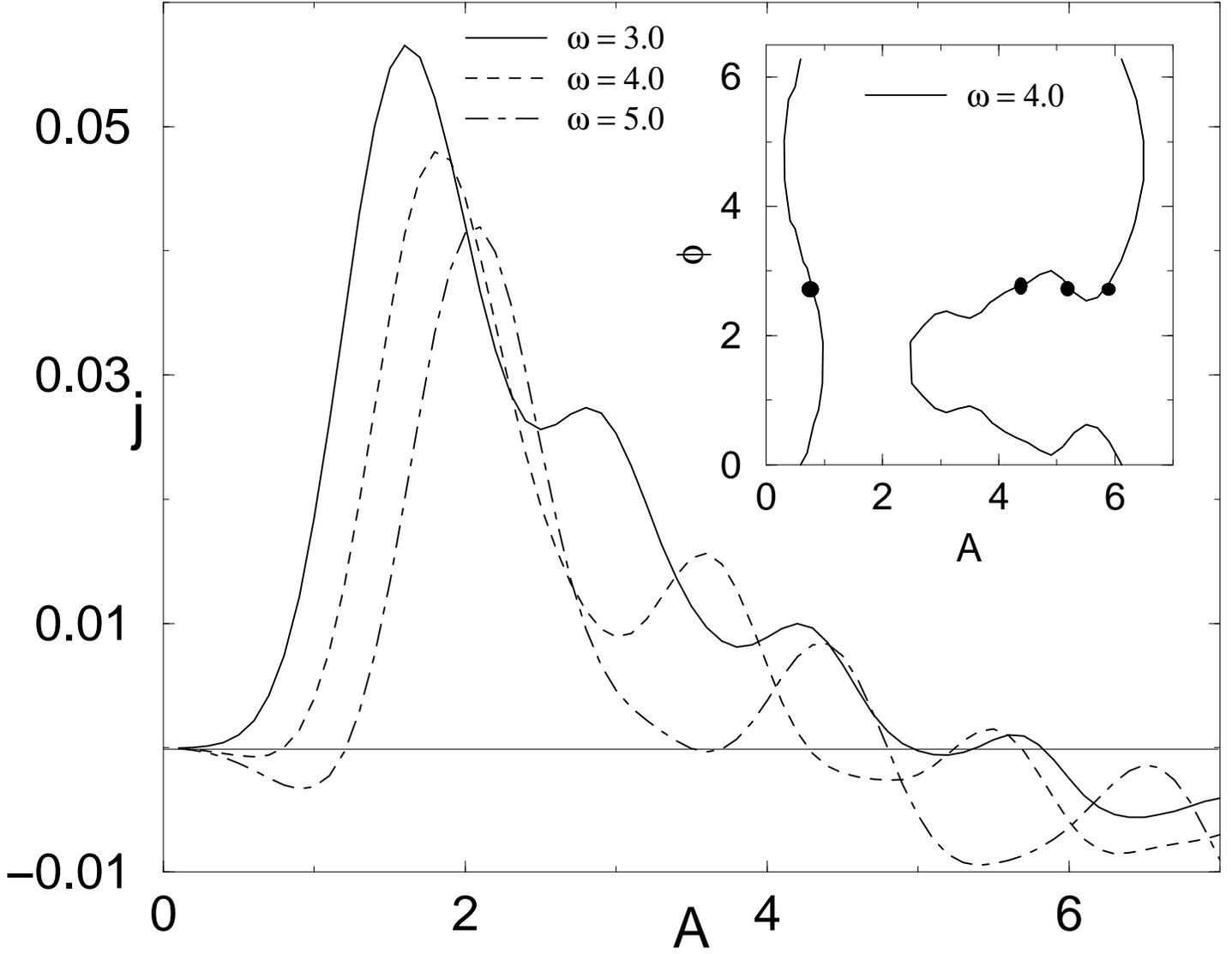}}
\vspace{.0in}
\caption{The mean current $j$ with amplitude $A$ of the forcing term
  for $\phi = 0.88 \pi,T=0.05$ and $\lambda=0.1$ with $w=3.0,
  4.0, 5.0$. The inset shows the contour of zero current for $T=0.05$
  for $w = 4.0$. The dots in the inset shows the four values of
  $A$ for $\phi = 0.9 \pi$ where the current reversal occurs. }
\label{j-A}
\end{figure}

\begin{figure}
\centerline{\epsfbox{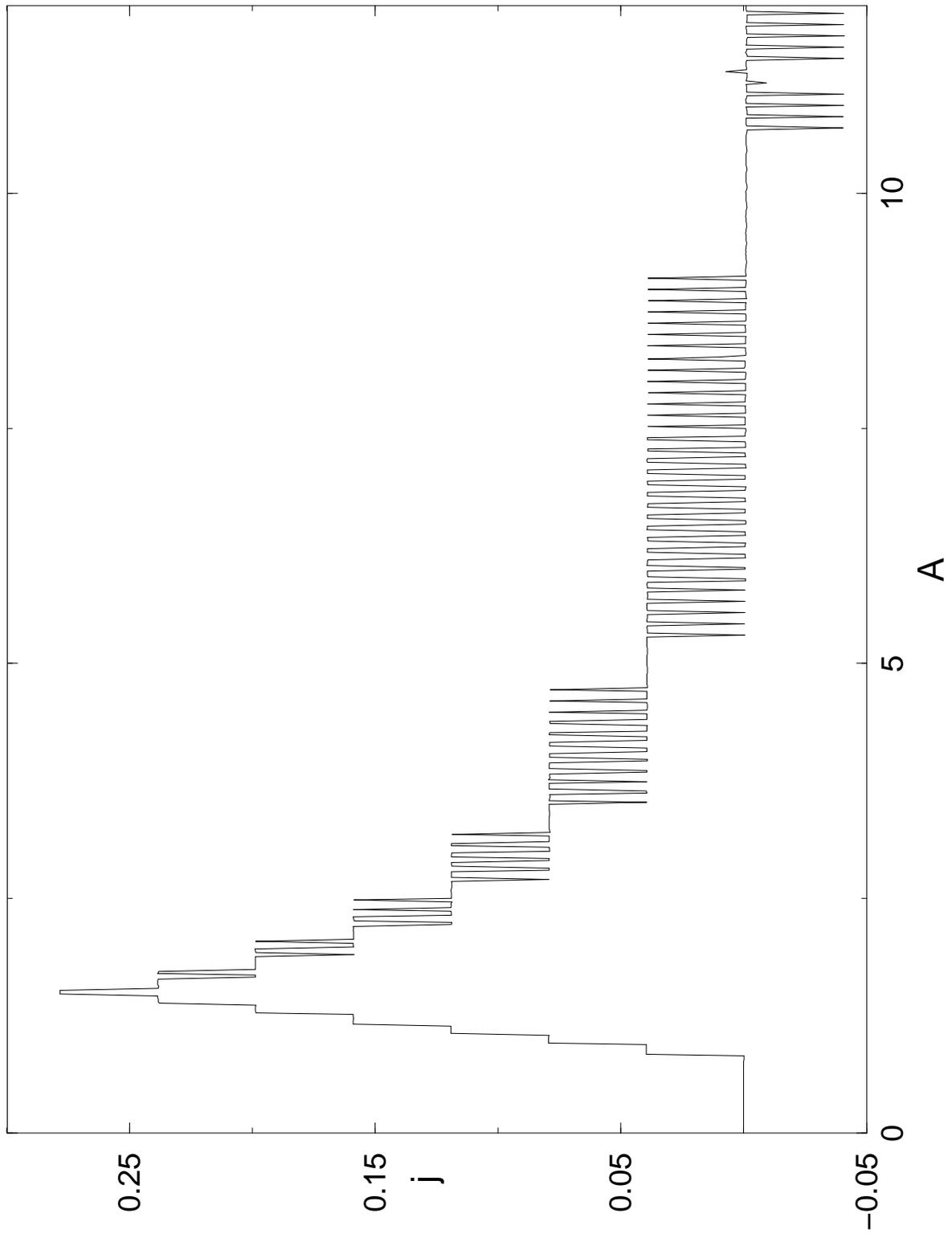}}
\vspace{.0in}
\caption{The deterministic current $j$ \textit{vs} amplitude of
  forcing $A$ for $w=0.25$ and $\phi=0.2 \pi$ and $\lambda = 0.1$.}
\label{Deter}
\end{figure}

Multiple current reversals can also be seen when the amplitude ( $A$) of the
forcing term is varied in a suitable parameter regime of our system.
In Fig.~\ref{j-A}, the plot of $j$ versus $A$ is shown for
different values of $w$, keeping $\lambda$, $\phi$ and $T$ fixed at
0.1, $0.88\pi$ and $0.05$ respectively. For $w = 4.0$ curve, we can see as
many as four current reversals. For very large value of $A$, the current
asymptotically goes to a constant value depending on the value of
$\phi$, as was previously shown for the adiabatic
case~\cite{Dan3}. This value was shown to be $-\frac{\lambda}{2} \sin
(\phi)$.  This is special to the space dependent friction
ratchet where the currents saturate to a finite value in the large $A$ 
limit. In the absence of space dependent friction it is to be
noted that currents decay to zero in the same asymptotic
regime. As the asymptotic value depends on the phase $\phi$, so we can 
choose it appropriately to make the asymptotic current positive or
negative. In the present case $\phi$ is chosen such that the
asymptotic current is negative which guarantees at least one current
reversal irrespective of frequency. The oscillatory behavior in the $j-A$
characteristics is the reminiscent of the deterministic dynamics
\cite{Main,Lich} which will be discussed later. The inset in Fig.~\ref{j-A} shows the
zero contour of $j$ versus $\phi$ and $A$ for $w = 4.0$. For $\phi > \pi$
only two current reversals can be seen. For $\pi > \phi > \pi - \epsilon$ 
(where $\epsilon$ is a small number ) or $0 < \phi < \epsilon$, four 
or more current reversals can occur. The value of $\epsilon$ depends 
critically on $T$, large for small $T$ and vice-versa. It is also to be noted
that we have even number of reversals for finite frequency driving. 
These oscillatory features along with their associated current reversals 
disappear in the high temperature regime as expected. 
Figure~\ref{Deter} is for $w = 0.25$ and shows that
current reversal is also possible
in the deterministic regime of the overdamped system. This deterministic
reversal of current cannot be attributed to the chaotic motion of the
system like in an forced underdamped oscillator~\cite{Under}. It solely arises due the 
presence of space dependent friction. As shown in Ref.~\cite{Main} here too 
the deterministic current shows  quantization and phase locking
behaviour. However all these complex features are not robust in  the 
presence of noise as discussed in earlier literature.

        In conclusion, we have studied the transport properties of
overdamped Brownian particles moving in an asymmetric potential 
 with space dependent 
friction coefficient and rocked by periodic force. We observe several novel
and complex features arising due to the interplay between asymmetry and 
inhomogeneous friction.
Currents in the low temperature regime is mostly influenced by the asymmetry
of the potential. At higher temperatures it is controlled by the modulation
parameter $\lambda$ of the friction coefficient. We find current reversal
with temperature even when the forcing is adiabatic. In the presence of 
finite frequency, twice current reversals are seen. As function of
amplitude of the forcing term we observe multiple current reversals.
Current even
reverses its sign in the adiabatic deterministic regime. 
All the above results can be understood in a qualitative manner.
We expect that our analysis should be applicable for the motion of particle
in porous media and for molecular motors where space dependent friction can
arise due to the confinement of particles.

   MCM acknowledges partial financial support and hospitality
from the Institute of Physics, Bhubaneswar. MCM and AMJ acknowledgepartial financial support from the Board of Research in Nuclear
Sciences, DAE, India.

%
%
%

\newpage

\end{document}